\documentclass[english,twoside,a4paper,10pt]{article}

\usepackage[latin1]{inputenc}
\usepackage[T1]{fontenc}
\usepackage{amsmath}
\usepackage{amsfonts}
\usepackage{graphicx}
\usepackage{subfigure}
\usepackage{a4wide}
\usepackage{amssymb}
\usepackage{fancyhdr}
\usepackage{mathrsfs}

\begin{document}

\title{\bf Static Spherically Symmetric Solutions in $F(R)$ Gravity }
\author{ 
L. Sebastiani\footnote{E-mail address:l.sebastiani@science.unitn.it
} and S. Zerbini\footnote{E-mail address:zerbini@science.unitn.it
}\\
\\
\begin{small}
Dipartimento di Fisica, Universit\`a di Trento, Italy and 
\end{small}\\
\begin{small}
 Gruppo Collegato di Trento, Istituto Nazionale di Fisica Nucleare, Sezione di Padova, Italy
\end{small}\\
}
\date{}

\maketitle


\begin{abstract}

A  Lagrangian derivation of the Equations of Motion (EOM) for static spherically symmetric metrics in $F(R)$ modified gravity is presented. For a large class of metrics, our approach permits to  reduce the EOM to a single equation and we show 
how it is possible to construct exact solutions in $F(R)$-gravity. All known exact solutions are recovered. We also 
exhibit a new  non trivial  solution with non constant Ricci scalar.

\end{abstract}



\section{Introduction}

Recent observational data imply that the current expansion of the universe
is accelerating. This is the so called Dark Energy issue.
There exist several descriptions of  this acceleration. Among them, the simplest one is the introduction of small positive 
Cosmological Constant in the framework of General Relativity (GR), the so called $\Lambda$-CDM model. 
A generalization of this simple modification of GR consists in considering  modified gravitational theories, 
in which the action is described by a function $F(R)$ of the Ricci scalar $R$ (for example, see~\cite{review6}-\cite{review7}). Typically these modified models admit the de Sitter (dS) space as a solution and the stability of this solution has been investigated in several places  (see, for example \cite{guido,guido2,Far,Monica}). Furthermore, viable $F(R)$ models, namely models able to pass the local gravitational GR tests as well as able to describe the inflation with dark energy in unified way, have been recently discussed \cite{Saw,seba10, Od1,Od2}.   

To begin with, we recall the Equations of Motion in vacuum for a general $F(R)$ model
\begin{equation}
G_{\mu\nu} \equiv R_{\mu\nu}-\frac{1}{2}Rg_{\mu\nu}= G^{{\mathrm{MG}}}_{\mu\nu}
 \,.\label{EqE}
\end{equation}
Here, $R_{\mu\nu}$ is the Ricci tensor and  the `modified gravity'  tensor
$G^{{\mathrm{MG}}}_{\mu\nu}$ is given by
\begin{equation}
G_{\mu\nu}^{{\mathrm{MG}}}=\frac{1}{F'(R)}\left\{\frac{1}{2}g_{\mu\nu}[F(R)-RF'(R)]
+(\nabla_{\mu}\nabla_{\nu}-g_{\mu\nu}\Box)F'(R)\right\}\,.
\end{equation}
The prime denotes derivative with respect to the curvature $R$,
$\nabla_{\mu}$
is the covariant derivative operator associated with $g_{\mu\nu}$ and
$\Box\phi\equiv g^{\mu\nu}\nabla_{\mu}\nabla_{\nu}\phi$ is the
d'Alembertian. The trace of Eq.(\ref{EqE}) reads
\begin{equation}
3\Box F'(R)+RF'(R)-2F(R)=0\,, \label{scalaroneq}
\end{equation}
which shows that there exists an additive  scalar dynamical degree of freedom represented by $F'(R)$, absent in GR.

The issue to find exact static spherically symmetric (SSS) solutions different from the Schwarzschild-dS one of these modified gravitational models appears a 
formidable task, since also for a reasonable $F(R)$, the equations of motions are much more complicated with respect of the 
ones in ``vacuum'', namely absence of matter, of General Relativity.  For the specific choice $R^{1+\delta}$, a class of 
exact SSS solution has been presented in \cite{CB}.

A  general discussion on SSS solutions based on above equation of motions has been presented in 
\cite{Multamaki,capo,bezerra, saffari}, where one can find further  references.

In this paper, we would like to present an alternative method for constructing SSS of a generic modified gravity models,
and, as a result, we also exhibit new exact SSS solutions.

The paper is organized as follows.
In Sec.~II, we present our Lagrangian derivation of the Equations of Motion for SSS metrics in $F(R)$-modified gravity. 
In Sec.~III, a large class  solutions are investigated. In Sec.~IV, we study the highly non trivial case of 
solutions with non constant scalar Ricci curvature.
Finally, conclusions are given in Sec.V.

We use units of $k_{B}=c=\hbar=1$ and denote the gravitational constant
$\kappa^2=8\pi G_N\equiv8\pi/M_{Pl}^2$ with the Planck mass of
$M_{PL}=G^{-1/2}_N=1.2\times 10^{19}\text{GeV}$.


\section{Lagrangian approach for static spherically symmetric vacuum solutions }

To begin with, we recall the action for modified $F(R)$-theories:
\begin{equation}
S=\frac{1}{2\kappa^2}\int d^4 x\sqrt{-g}F(R)\,,\label{action} 
\end{equation}
where $g$ is the determinant of metric tensor $g_{\mu\nu}$ and $F(R)$ is a generic function of the Ricci scalar $R$.  

We shall look for static spherically symmetric (SSS) solutions  of the type,
\begin{equation}
ds^2=-B(r)\mathrm{e}^{2\alpha(r)}dt^2+\frac{dr^2}{B(r)}+r^2d\Omega\,,
\end{equation}
where $d\Omega=r^2(d\theta^2+\sin^2\theta d\phi)$ and $\alpha(r)$ and $B(r)$
are unknown functions of $r$. With this ansatz, the scalar curvature reads
\begin{eqnarray}
R  &=&
-3\,\left[{\frac{d}{dr}}B\left(r\right)\right]{\frac{d}{dr}}
\alpha\left(r\right)-2\,B\left(r\right)\left[{\frac{d}{dr}}
\alpha\left(r\right)\right]^{2}-{\frac{d^{2}}{d{r}^{2}}}
B\left(r\right)-2\,B\left(r\right){\frac{d^{2}}{d{r}^{2}}}\alpha\left(r\right)\nonumber\\
&&-4\,{\frac{{\frac{d}{dr}}B\left(r\right)}{r}}
-4\,{\frac{B\left(r\right){\frac{d}{dr}}\alpha\left(r\right)}{r}}-2\,{\frac{B\left(r\right)}{{r}^{2}}}
+\frac{2}{{r}^{2}}\,.\label{R}
\end{eqnarray}
By plugging this expression into the action (\ref{action}), one obtains a higher derivative Lagrangian theory. 
In order to work with a first derivatives Lagrangian system, we may use the  method of Lagrangian multipliers, used  
for the FRW space-time in  Refs.\cite{vile,Capozziello,Monica}. In the static case we are dealing with, the method  
permits to consider as independent Lagrangian coordinates 
the scalar curvature $R$, and the quantities $\alpha$ and $B$, appearing in the spherically static symmetric ansatz. 
As a consequence, we will obtain two equations of motion where one of the unknown quantities $\alpha$ appears in a very simple way. The main difference with respect to the other general appraoches is that we do not directly make use of EOM 
(\ref{EqE}).   

By introducing the Lagrangian multipliers $\lambda$ and making use of Eq. (\ref{R}), the  action  (\ref{action}) may be written 
\begin{eqnarray}
S &\equiv& \frac{1}{2\kappa^2}\int dt\int d{ r}\left(e^{\alpha(r)}r^2 \right)\left\{ F(R)-\lambda \left\{R+\left\{3\,\left[{\frac{d}{dr}}B\left(r\right)\right]{\frac{d}{dr}}
\alpha\left(r\right)\right.\right. \right.    \nonumber\\
&& +2\,B\left(r\right)\left[{\frac{d}{dr}}
\alpha\left(r\right)\right]^{2}+{\frac{d^{2}}{d{r}^{2}}}
B\left(r\right)+2\,B\left(r\right){\frac{d^{2}}{d{r}^{2}}}\alpha\left(r\right)+4\,{\frac{{\frac{d}{dr}}B\left(r\right)}{r}}\nonumber\\ 
&&+4\,{\frac{B\left(r\right){\frac{d}{dr}}\alpha\left(r\right)}{r}}+2\,{\frac{B\left(r\right)}{{r}^{2}}}
\left.\left.\left.-\frac{2}{{r}^{2}}\right\}\right\}\right\}\,.
\end{eqnarray}
 Making the variation with respect to $R$, one gets
\begin{equation}
 \lambda=F'(R)\,.
\end{equation}
The prime denotes the derivative with respect to the curvature $R$. Thus, by substituting this value and by making an integration by part, the Lagrangian takes the form
\begin{eqnarray}
L(\alpha, d\alpha/dr, B, d B/dr, R, d R/dr)&=&e^{\alpha}\left\{r^2\left(F(R)-F'(R)R\right)+2F'(R)\left(1-r\frac{d B(r)}{dr}- B(r)\right)\right.
\nonumber\\
&& +\left.F''(R)\frac{d R}{d r}r^2\left(\frac{d B(r)}{d r}+2B(r)\frac{d \alpha(r)}{dr}\right)\right\}\,.
\end{eqnarray}
Making the variation with respect to $\alpha$, one gets the first equation of motion  
\begin{eqnarray}
\label{one}& &\frac{RF'(R)-F(R)}{F'(R)}-2\frac{\left(1-B(r)-r(dB(r)/dr)\right)}{r^2}\\\nonumber
& & +\frac{2B(r)F''(R)}{F'(R)}\left[\frac{d^2 R}{d
r^2}+\left(\frac{2}{r}+\frac{dB(r)/dr}{2 B(r)}\right)\frac{d R }{d
r}+\frac{F'''(R)}{F''(R)}\left(\frac{d R}{d
r}\right)^2\right]=0\,.
\end{eqnarray}
The variation with respect to $B(r)$ leads the second equation of motion
\begin{equation}
\left[\frac{d\alpha(r)}{dr}\left(\frac{2}{r}+\frac{F''(R)}{F'(R)}\frac{d
R}{d r}\right)-\frac{F''(R)}{F'(R)}\frac{d^2 R}{d
r^2}-\frac{F'''(R)}{F'(R)}\left(\frac{d R}{d
r}\right)^2\right]=0\,,\label{two}
\end{equation}
while by making the variation with respect to $R$, we recover the Eq.(\ref{R}). 

Once $F(R)$ is given, together with 
equation (\ref{R}), the above equations form  a system of three differential equations in the three unknown 
quantities $\alpha(r), B(r)$ and $R(r)$. We would like to 
note that one advantage of this approach is that the  $\alpha$ does not appear in Eq.(\ref{one}) and vice versa.
In the next Sections, we will find exact solutions of the above system of differential equations related to specific choices 
for $F(R)$. 

As first check of the formalism, let us look for the well known case of constant curvature solutions $R=R_0$.  From Eq.(\ref{two}) we have 
$\alpha=Const$ and it is easy to show that the only solution of Eqs.(\ref{R}) and(\ref{one}) is the Schwarzshild-De Sitter 
solution
\begin{equation}
B(r)=1-\frac{\Lambda}{3}r^2+\frac{C}{r}\,, 
\end{equation}
where $C$ is a generic constant and 
\begin{equation}
 \Lambda=\frac{1}{2}\left(R_{0}-\frac{F(R_{0})}{F'(R_{0})}\right)\,,
\end{equation}
with $R_0=4\Lambda$.

\section{Solution with constant $\alpha$}

Now, let us consider the case of non constant Ricci curvature, but still with $\alpha=Const$.  From Eq.(\ref{two}):
\begin{equation}
F'''\left(\frac{d R}{dr}\right)^2+F''\left(\frac{d^2 R}{d r^2}\right)=0=\frac{d^2}{dr^2}F'(R)\,. 
\end{equation}
Thus,
\begin{equation}
F'(R)=ar+b\,. \label{Fprime}
\end{equation}
Here, $a$ and $b$ are two integration constants. If we give the explicit form of $R$, we may  find $r$ as a function 
of Ricci scalar and reconstruct $F'(R)$ realizing such solution.
The equation (\ref{R}) leads to:
\begin{equation}
R  =
-{\frac{d^{2}}{d{r}^{2}}}
B\left(r\right)
-4\,{\frac{{\frac{d}{dr}}B\left(r\right)}{r}}
-2\,{\frac{B\left(r\right)}{{r}^{2}}}
+\frac{2}{{r}^{2}}\,.\label{R2}
\end{equation}

Since $(F''(R))d R/dr=d F'(R)/dr=a$ and $d F(R)/d r=F'(R) d R/d r$, by multiplying Eq.(\ref{one}) with respect to $F'(R)$ and by deriving with respect to $r$, one has
\begin{equation}
-\frac{d^2 B(r)}{dr^2}\left(a+\frac{b}{r}\right)+\frac{2 a}{r^2}\left(2B(r)-1\right)+\frac{2b}{r^3}(B(r)-1)-\frac{a}{r}\frac{d B(r)}{d r}=0\,.\label{equazione} 
\end{equation}

If $a=0$, the solution of this equation is the Schwarzschild-de Sitter solution, which corresponds to $F'(R)=b$, namely
GR plus a cosmological constant $F(R)=bR+F_0$. 
If $b=0$, the general solution is
\begin{equation}
B(r)=\frac{1}{2}\left(1-\frac{C_{1}}{r^2}+C_2r^2\right)\,. 
\label{L1}
\end{equation}
Eq.(\ref{R2}) leads to
\begin{equation}
 r=\sqrt{\frac{1}{R+6C_2}}\,,
\end{equation}
so that
\begin{equation}
 F'(R)=a\sqrt{\frac{1}{R+6C_2}}\,.
\end{equation}
We then can reconstruct $F(R)$ and we get the model
\begin{equation}
 F(R)=\frac{a}{2}\sqrt{R+6C_2}\,,\label{L0}
\end{equation}
for which Eq. (\ref{L1}) is a spherically symmetric static solution with $\alpha=0$. If we put $C_2=0$, we recover a special case of Clifton-Barrow solution \cite{CB}. If $C_1=0$, we recover a special case reported in \cite{Multamaki}.

The most general solution of Eq.(\ref{equazione}) for $b \neq 0$ is
\begin{eqnarray}
\label{generic}B(r)&=& \left\{-\frac{r^2 C_2\log\left(r\right)a^3}{b^4}+\frac{r^2 C_2\log\left(b+a r\right)a^3}{b^4}+\frac{3 r^2 a^2}{2 b^2}-\frac{r C_2 a^2}{b^3}-\frac{r^2\log\left(r\right)a^2}{b^2}\right.\\\nonumber
&&+\frac{r^2\log\left(b+a r\right)a^2}{b^2}-\frac{r a}{b}+\frac{C_2 a}{2 b^2}+r^2 C_1\left.-\frac{C_2}{3 b r}+1\right\}\,,
\end{eqnarray}
where $C_1$ and $C_2$ are generic constants. For $a=0$, one again obtains the Schwarzschild-De Sitter  solution. 
The above solution is one of the main result of the paper, because it is the starting point of the reconstruction method, and it is compatible with the result obtained in Ref. \cite{saffari}. 
Implicitly $F(R)$ is determined by Eqs. (\ref{Fprime}),  (\ref{R2}) and  (\ref{generic}). If a $F(R)$-model realizes the metric
\begin{equation}
ds^2=-B(r)dt^2+\frac{dr^2}{B(r)}+r^2d\Omega^2\,, 
\end{equation}
the coefficient $B(r)$ assumes the generic form of Eq.(\ref{generic}).

On the other hand, we have to note that, since Eq.(\ref{equazione}) has been obtained trough a derivation, the left side of Eq.(\ref{one}) evaluated on the solution (\ref{generic}) could be proportional to  $1/F'(R)$, so that some other constraints on free parameters could be necessary. As an example, let us consider the simple case where
\begin{equation}
C_2=-\frac{b^2}{a}\,. 
\end{equation}
Since $b$ is adimensional, we assume $b=1$ and Eq.(\ref{generic}) becomes
\begin{equation}
B(r)=\frac{1}{2}+\frac{1}{3ar}+\frac{3a^2r^2}{2}+C_{1}r^2\,, 
\end{equation}
and the Ricci scalar reads
\begin{equation}
R=\frac{1}{r^2}-12C_1-18a^2\,. 
\end{equation}
By using Eq.(\ref{Fprime}) one has
\begin{equation}
F(R)=R+2a\sqrt{R+18a^2+12C_{1}}\,. 
\end{equation}
Now, we find that the left side of Eq.(\ref{one}) is $3(3a^2+2C_1)/F'(R)$, so that we have to require $C_1=-3a^2/2$. As a result, the reconstruction gives
\begin{equation}
F(R)=R+2a\sqrt{R}\,,\label{LL} 
\end{equation}
and this model  admits a spherically symmetric solution with constant $\alpha$ parameter and 
\begin{equation}
B(r)=\frac{1}{2}\left(1+\frac{2}{3ar}\right)\,, 
\label{L}
\end{equation}
 and Ricci curvature $R=1/r^2$.

\section{Solution with non constant curvature}

In this Section, we will look for  solutions with non constant $\alpha$ and non constant Ricci scalar. Thus, we 
make the ansatz
\begin{equation}
\alpha=\frac{1}{2}\log\left(\frac{r}{r_0}\right)^q\,,
\end{equation}
\begin{equation}
R=\delta r^s\,,\label{Rform} 
\end{equation}
where $r_0, q, \delta$ and $s$ are constant parameters. From Eq.(\ref{two}), it follows that the modified  gravity model 
which realizes this kind of solution is
\begin{equation}
F(R)=k(R^{\gamma}+\mu)\,, 
\end{equation}
where $k$ and $\mu$ are  constants and $\gamma$ is given by 
\begin{equation}
 \gamma=\frac{(2+q+4s)s\pm s\sqrt{q^2+20q+4}}{4s^2}\,.
\end{equation}
If we put $s=-2$, one has $q=2(\gamma-1)(2\gamma-1)/(2-\gamma)$. In this case, Eq.(\ref{one}) has a simple solution by choosing $\mu=0$ and $\delta=6\gamma(\gamma-1)/(2\gamma^2-2\gamma-1)$. Thus, one recovers the Clifton-Barrow solution \cite{CB}
\begin{equation}
 B(r)=\frac{(\gamma-2)^2}{(10\gamma-4\gamma^2-7)(2\gamma^2-2\gamma-1)}\left(1-\frac{C}{r^{(4\gamma^2-10\gamma+7)/(2-\gamma)}}\right)\,,
\end{equation}
which is consistent with Eq.(\ref{R}), compatibly with assumption (\ref{Rform}).\\
\\
However, there exists also  another solution with $q=1$ and $s=-1$. It means, $\gamma=-1$ or $\gamma=3/2$. 
Eq.(\ref{R}) is solved by
\begin{equation}
B(r)=\frac{4}{7}-\frac{2}{3\sqrt{-6\mu}}r+\frac{C_{1}}{r^{7/2}}+\frac{C_{2}}{r}\,. 
\end{equation}
Here, $C_{1}$ and $C_{2}$ are arbitrary constants. On the other hand, it is easy to see that Eq.(\ref{one}) is inconsistent if $\gamma=3/2$, but admits a solution for $C_2=0$ if $\gamma=-1$ and
\begin{equation}
 \delta=\sqrt{-\frac{6}{\mu}}\,,
\end{equation}
with $\mu$ negative. As a consequence, we may put $\mu=-\frac{h^2}{6}$, $h>0$ and we have found that the model
\begin{equation}
F(R)=k\left(\frac{1}{R}-\frac{h^2}{6}\right)\,, 
\end{equation}
 admits the following SSS  metric 
\begin{equation}
ds^2=-\left(\frac{r}{r_0}\right)B(r)d t^2+\frac{d r^2}{B(r)}+r^2d\Omega\,, 
\end{equation}
with
\begin{equation}
B(r)=\frac{4}{7}\left(1-\frac{7}{6 h}r+\frac{C}{r^{7/2}}\right)\,. 
\label{N}
\end{equation}
The Ricci scalar becomes
\begin{equation}
 R=\frac{6}{h\,r}\,.
\end{equation}

\section{Concluding remarks}

In this paper, a new method for investigating SSS solutions for a generic $F(R)$ has been presented.  
It is our opinion that our method is quite general and offers some advantages with respect the other  general methods
presented in \cite{Multamaki,capo,bezerra,saffari}. In fact in the important case of  $\alpha$ constant 
(in this case, we may put  $\alpha=0$), we obtain a general form of the metric via the explicit expression for the quantity $B(r)$. 

The static solutions describe a black hole as soon as there exists a real positive solution of
$B(r_H)=0$. If this happens, there exists an event horizon and the related  Hawking temperature reads
\begin{equation}  
T_H=\frac{1}{4\pi}\frac{d B(r_H)}{dr}\,.
\end{equation}
This is well known result, and it can be justified in several ways, like the elimination of conical singularity in the 
associated Euclidean metric, or making use of the tunneling method introduced by Parikh and Wilczek \cite{PW,A}.

For example, for the model $F(R)=\sqrt{R+6C_2}$ of Eq.(\ref{L0}), since the positive solutions of $B(r_H)=0$ read
\begin{equation}  
r_H=\sqrt{\frac{-1\pm\sqrt{1+4C_1C_2}}{2C_2}}\,,
\end{equation}
one has to require $C_1>0$ and $C_2>0$ (for the plus sign) or $C_2<0$ and $C_1<-1/(4C_2)$ (for the sign minus). 

For the other model in Eq.(\ref{LL}), we have to chose $a=-g<0$, $g>0$, thus $F(R)=R-2g\sqrt{R}$ and
we have
\begin{equation}  
r_H=\frac{2}{3g}\,.
\end{equation}
With regard to the other class of solutions we have discussed, the ones with non trivial $\alpha(r)$, the situation is 
different. A discussion concerning the Clifton-Barrow class has been presented elsewhere\cite{Bel}. Thus we consider only 
the new solution (\ref{N}) corresponding to the model $F(R)=k\left (\frac{1}{R}-\frac{h^2}{6}\right)$. 
In this case, if $C>0$, there exists
a positive root of 
\begin{equation}  
1-\frac{7}{6h}r+\frac{C}{r^{7/2}}=0\,,
\end{equation}
since  $d B(r)/dr<0$, namely is non vanishing. Thus, as in de Sitter space-time, we may take the absolute value
of the surface gravity for the expression of Hawking temperature. Furthermore, besides the usual Killing vector associated 
with the static space-time, on general ground, it is also possible to introduce the Kodama vector, with an associated 
surface gravity, the Hayward's surface gravity. A detailed discussion about this issue can be found in 
\cite{noi1,noi2,noi3}. As far as the other thermodynamical quantities associated wit these black holes solutions, 
the entropy can be calculated by the Wald method \cite{W,V,F}, and one has \cite{guido}
\begin{equation}  
S_H=\frac{A_H}{4}F'(R_H)\,.
\end{equation} 
In the last example, we may choose the dimensional constant $k$ negative in order to deal with a positive entropy.

Much more difficult is the issue associated with  the energy or mass of these black hole solutions. 
With regard to this, we refer to the papers \cite{D,gong,V1,Cai}. 

Finally we note that the functional freedom in the choice of $F(R)$ gives the possibility of its cosmological reconstruction as it has been reviewed in \cite{review6}. In fact, following the construction developed in this paper and BH reconstruction outlined in 
Ref. \cite{Od3}, we can generate more black holes in F(R) gravity via its reconstruction.

\end{document}